\documentclass[aps,prb,preprint,showpacs,superscriptaddress,
preprintnumbers,amsmath,amssymb]{revtex4}

\usepackage{graphicx}
\usepackage{dcolumn}
\usepackage{bm}
\usepackage{ifthen}
\usepackage{booktabs}
\def\mc{\multicolumn}

\begin{document}

\title{Systematically improvable optimized atomic basis sets for {\it ab inito} calculations}
\author{Mohan Chen}
\affiliation{Key Laboratory of Quantum Information, University of
Science and Technology of China, Hefei, 230026, People's Republic of
China}
\author{G-C Guo}
\affiliation{Key Laboratory of Quantum Information, University of
Science and Technology of China, Hefei, 230026, People's Republic of
China}
\author{Lixin He \footnote{corresponding author: helx@ustc.edu.cn} }
\affiliation{Key Laboratory of Quantum Information, University of
Science and Technology
of China, Hefei, 230026, People's Republic of China}
\date{\today }

\begin{abstract}
We propose a unique scheme to construct fully optimized atomic basis
sets for density-functional calculations. The shapes of the radial
functions are optimized by minimizing the {\it spillage} of
the wave functions between the atomic orbital calculations and
the converged plane wave calculations for dimer systems. The quality
of the bases can be systematically improved by
increasing the size of the bases within the same
framework. The scheme is easy to implement and very flexible. We
have done extensive tests of this scheme for wide variety of
systems. The results show that the obtained atomic basis sets
are very satisfactory for both accuracy and transferability.
\end{abstract}


\maketitle

\section{Introduction}

The last few years have seen the development of the first-principles
methods in the application of complex material systems containing
hundreds and thousands of atoms.
\cite{otsuka08,volker08,skylaris08,artacho08} This is made possible
because of the so called linear scaling algorithms
\cite{goedecker99} are used, by taking the advantages of the
locality of the electronic structures.\cite{kohn96}
The widely used plane wave basis is not
suitable for the linear scaling algorithms, because of its extended
nature. Instead, the local basis, such as atomic orbitals are the
better choices.

The advantage of atomic orbitals are two folds. First, the basis
size of atomic orbitals is much smaller compared to other
basis sets, such as plane wave, and real space mesh, etc. Second,
the atomic orbitals are strictly localized and therefore compatible
with the linear scaling algorithms \cite{goedecker99} for electronic
calculations.
However, the atomic basis sets must be constructed
very carefully to ensure both good accuracy and
transferability. Furthermore the quality of the basis sets
should be systematically improvable in an unbiased way.

Several schemes to construct atomic orbitals have been
proposed.\cite{soler02,ozaki03,volker08} For example, the atomic
orbitals can be constructed by applying certain confinement
potentials to the isolated atoms. \cite{sankey89,junquera01} To
ensure the transferability of the orbitals, one could use larger
basis set, by using more than one radial function for each angular
moment (multi-$\zeta$), or by including higher angular moment
orbitals (polar orbitals). Empirically, the multi-$\zeta$ functions
can be generated by split-valence method, \cite{soler02} whereas the
polar orbitals can be generated by applying electric field in
addition to the confinement potential. These methods have been
demonstrated to be effective. \cite{soler02} Nevertheless, different
level of orbitals are treated in different ways, and are not
guaranteed to be the optimized ones. \cite{artacho08}
Volker {\it et al.} proposed a way to systematically improve the
basis that they choose the one that improves the energy most from a
large pool of {\it pre-selected} orbitals.\cite{volker08}
Alternatively, Ozaki optimize shape of atomic orbitals adopted to
different environment as part of the self-consistency
cycle.\cite{ozaki03} However, in this scheme, every atom must has
different orbital shape even for the same element.

In this work, we propose a unique method that allows to construct
systematically improvable fully optimized atomic basis sets for
density-functional calculations. Unlike previous methods, all the
orbitals (including multi-$\zeta$ and polar orbitals) can be
constructed in a same procedure. The shapes of the radial functions
are optimized by minimizing the {\it spillage} of the wave functions
between the converged plane wave calculations and those from atomic
orbital calculations for dimer systems, and therefore no
pre-assumption about the radial functions is needed.
The scheme is easy to implement and is very flexible and efficient.
We have done extensive tests of this scheme for wide variety of
systems. The results are very promising, showing very satisfactory
results for both accuracy and transferability.

The rest of paper is organized as follows. In Sec.~\ref{sec:methods}
we give detailed introduction of our scheme to construct atomic
orbitals. We test the obtained orbitals by calculating the
structural and electronic properties of wide variety of systems in
Sec.~\ref{sec:results}, including III-V and group IV semiconductors,
and GaN, ZnO, Al, Pb and MgO etc. We summarize in
Sec.~\ref{sec:conclusion}.

\section{Methods}
\label{sec:methods}


One of the most popular ways to generate the atomic orbitals is to
use atomic orbitals of isolated atoms in certain confinement
potentials. This procedure usually gives the minima basis of the
atom. To make the basis more complete, one has to use multi-zeta
orbitals and polar orbitals. The multi-zeta
orbitals can be obtained via a split-valence
method, \cite{soler02} whereas the polar orbitals are generated by
applying constant electric fields. \cite{soler02} Obviously, the
orbitals are constructed in very different
procedures. The quality of the orbitals are uncontrolled, even
though they are usually good. When larger basis sets are
needed for high quality calculations, the procedures to get the
orbitals are tedious.

We use a very different strategy to construct fully optimized atomic
orbitals that are highly transferable. The strategy is based on
minimizing the spillage of the wave functions between the atomic
orbital calculations and the plane wave results.
The spillage is a measurement of the difference between the Hilbert
space spanned by a set of local basis and the space spanned by the
``exact'' wave functions of the interested states of given
systems.\cite{portal95,portal96} The spillage is defined as,
\cite{portal95}
\begin{equation}
\mathcal{S}=\frac{1}{N_{n}}\sum_{n=1}^{N_{n}}\langle \Psi_{n} | 1 -
\hat{P} | \Psi_{n} \rangle \, ,
\end{equation}
where $\Psi_{n}$ is the plane wave calculated eigenstate and $N_{n}$
is the number of states of interest. $\hat{P}$ is a projector which
spanned by all the atomic orbitals, i.e.,
\begin{equation}
\hat{P}=\sum_{\mu\nu}|\phi_{\mu}\rangle S_{\mu\nu}^{-1}
\langle \phi_{\nu}| \, ,
\end{equation}
where $\phi_{\mu}=\phi_{\mu}({\bf r} -{\bf r}_{\mu})$ is the
$\mu$-th local orbital. $S_{\mu\nu}$ is the overlap matrix between
orbitals $\phi_{\mu}$ and $\phi_{\nu}$, i.e.,
\begin{equation}
S_{\mu\nu}=\langle\phi_{\mu}|\phi_{\nu}\rangle \, ,
\end{equation}
where $\mu = \{\alpha, i, \zeta, l, m\}$, in which $\alpha$ is the
element type, $i$ is the index of atom of each element type,
$\zeta$ is the multiplicity of the radial functions for the angular
momentum $l$, and $m$ is the magnetic quantum number.

The spillage has been applied to analysize the quality of given
atomic basis sets. \cite{portal95,portal96,kenny00,gusso08} There
are also some attempts to choose optimized local basis for a given
system by minimizing the spillage value.
\cite{portal95,portal96,kenny00} In these methods, it usually starts
from certain pre-assumed orbital shapes with few free parameters.
The spillage is then used to determine these parameters. In Ref.
\onlinecite{portal96}, the authors used the spillage to choose the
best Slater-type orbitals or the pseudo atomic orbitals for a given
system. However, the transferability of the atomic orbitals are not
taken into consideration. Using similar idea, Kenny {\it et al.}
took one step further and generated multi-$\zeta$ and polar
orbitals.\cite{kenny00}


Here we propose a new scheme based on the spillage formulism to
generate high quality atomic orbitals that are systematically
improvable. Our method improves upon that of previous methods in three
aspects: (i) The shape of atomic orbitals can be generated
automatically without any pre-assumptions. (ii) The atomic basis can be
systematically improved within the same framework.
(iii)The transferability of the atomic bases is improved by
carefully choosing the reference systems.

\subsubsection{The radial functions}

In our scheme, each atomic orbital is written as a radial function
multiplied by a spherical harmonic function. The radial functions
are expanded into spherical Bessel functions. The $\mu$-th local
orbital is
$\phi_{\mu}(\mathbf{r})=f_{\mu,l}(\mathbf{r})Y_{lm}(\hat{r})$,
where,
\begin{equation}
f_{\mu,l}(\mathbf{r})=\left\{
\begin{array}{ll}
\sum_{q}c_{\mu q}j_l(qr), & r < r_c\\
0 & r \geq r_c \, .\\
\end{array}
\right.
\end{equation}
$j_l(qr)$ is the spherical Bessel function. $q$ is chosen to
satisfy $j_{l}(qr_c)$=0, where $r_c$ is the cut off radius of the
radial functions.
The atomic orbitals are therefore strictly zero beyond $r_c$.
$Y_{lm}(\hat{r})$ is the spherical harmonic
function, in which $l$ is the angular momentum, $m$ is the magnetic
quantum number. The coefficients $c_{\mu q}$ are chosen to minimize
the spillage of the reference systems via a simulated annealing
method. Since the first-principles total energy calculation can be
done once for all for the reference system, the method is very
efficient.

We use the same energy cutoff of plane wave calculation for the
spherical Bessel functions. For pseudopotentials calculations, 15
$\sim$ 30 spherical Bessel functions are usually good enough to
obtain reliable atomic orbitals.

In order to make the kinetic energy integral well defined, one needs
to make the second derivative of the atomic orbitals continuous.
This can be done by multiplying the radial part of the atomic
orbitals by a smooth function,\cite{kenny00}
\begin{equation}
g(r)=1-\exp[-\frac{(r-r_{\rm cut})^{2}}{2\sigma^{2}}]\, .
\end{equation}
In our test, we find $\sigma$  has little influence on the final
results, we thus fix $\sigma$=0.1.

In our scheme, we do not have to assume the shape of the atomic
orbitals, therefore in principle we can get the fully optimized radial
functions.

\subsubsection{Reference systems}

One first needs to generate the atomic basis sets for some reference
systems, then use it to more general cases, assuming the atomic
basis sets are transferable.
A good reference system is therefore crucial for generating high
quality transferable basis set. Here, we choose dimer systems as
reference systems. The spillage of the system is defined to be the
average spillage values of the few selected dimers.\cite{volker08}
We choose several dimers at different bonding lengths, covering the
whole dissociate curve of the dimer. If only one dimer is used as
the reference, it will leave a finger print into the atomic
orbitals, therefore worsen their transferability. We find 4 or 5
dimers are enough to generate reliable local orbitals. Further
increasing the number of dimers to the reference system as many as
20 does not significantly improve the results. As it will be shown
in the paper, the atomic basis generated from the dimer systems can
be used to calculate different bulk systems with high accuracy,
showing remarkable transferability. This is very important for
studying complex material system, which may have complex chemical
environment in a single system, including defects, surfaces, alloy
etc.

\subsubsection{Systematically generate atomic orbitals}

The quality of the numerical orbital basis set can be systematically
improved by increasing the number of radial functions (multi-zeta
orbitals) of given angular momentum and by including orbitals with
higher angular momentum (polar orbitals). There are several ways in
the literature to construct multi-zeta orbitals and polar orbitals.
However, there is still no systematically way to generate multi-zeta
orbitals and polar orbitals, in which all orbitals are treated in a
unbiased way. \cite{artacho08} In contrast, in our scheme all
orbitals can be generated with the same procedures. To do so, we
first generate the orbitals with given angular momentum, which we
can call level 1 orbitals. The higher level orbitals can be
generated using the same procedure, by minimize the spillage of the
remaining Hilbert space, which orthogonal to the space spanned by
the all previously generated atomic orbitals.
Taken Si DZP (i.e., double $\zeta$ functions plus one polar orbitals
basis as an example. In the first step, we generate
the first $s$ and $p$ orbitals, which form a minimal basis set for
Si. In the second step, we generate the second $s$ and $p$ orbitals
(multi-zeta orbitals).
We first orthogonalize the wave functions of the reference states to
the atomic orbitals generated in step 1. Here we define the
projector operator formed by level 1 orbitals as,
\begin{equation}
\hat{P}^{(1)}=\sum_{\mu\nu}|\phi_{\mu}^{(1)}\rangle S_{\mu\nu}^{-1}
\langle \phi_{\nu}^{(1)}|
\end{equation}
where $\phi_{\mu}^{(1)}$ is the $\mu$-th orbital of level 1 orbitals. The
remaining wave functions are,
\begin{equation}
|\Psi_{n}^{(2)}\rangle = (1 - \hat{P}^{(1)}) |\Psi_{n}^{(1)} \rangle
\label{eqs:projection}
\end{equation}
where $|\Psi_{n}^{(2)}\rangle$ is the new set of wave functions,
which is orthogonal to the atomic orbitals generated at step 1,
i.e.,
\begin{equation}
\hat{P}^{(1)}|\Psi_{n}^{(2)}\rangle=0 \, .
\end{equation}
We then minimize the spillage between the second $s$, $p$ orbitals
and the space spanned by $|\Psi_{n}^{(2)}\rangle$. In step 3, we
generate the $d$ orbitals following exactly the same procedures.
The order of the orbitlas added into the basis can be determined by choosing
the orbitals that decrease the spillage most after the orbitals have
been added to the basis.
In such way, we can systemtically generate orbitals of
any multiplicity and angular momentum in a unified scheme.
This is important if high accuracy
calculations are needed.

\subsubsection{Optimize the shape}

\begin{figure}
\centering
\includegraphics[width=3in]{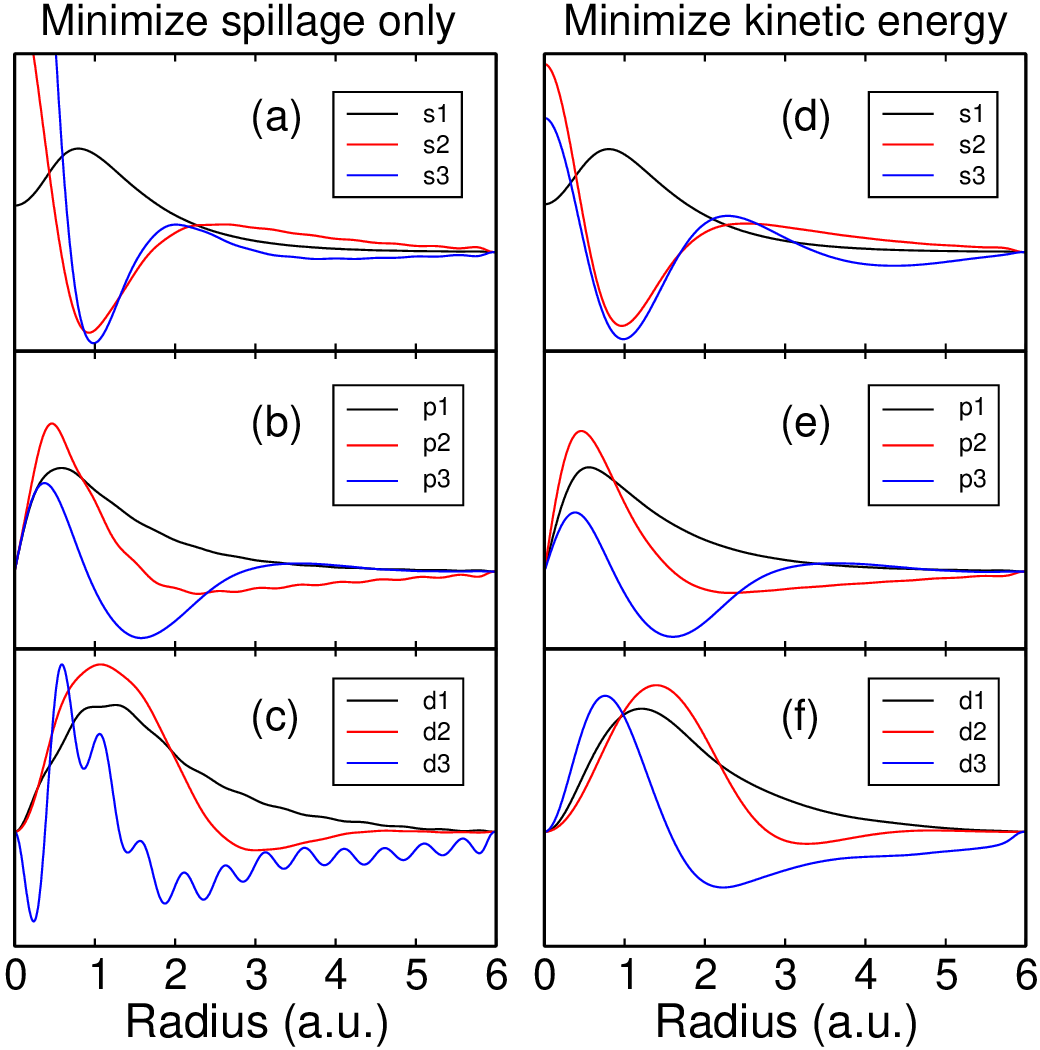}
\caption{(Color online) (a), (b), (c) The radial functions of unoptimized carbon
$s$, $p$, $d$ orbitals respectively. (d), (e), (f) The radial
functions of optimized carbon $s$, $p$, $d$ orbitals.}
\label{fig:shape}
\end{figure}

To give an idea of how the obtained radial functions look like using
the above scheme, we show the radial functions of the first 3  $s$,
$p$, $d$ atomic orbitals in Fig. \ref{fig:shape} (a), (b), (c)
respectively for the carbon atom. The orbitals are obtained by
taking five carbon dimers of different bond lengths as reference
systems. The bond length of the dimmers are chosen to be 1.00, 1.25,
1.50, 2.00, 2.50 \AA. The energy cutoff of plane wave basis
calculations is set to 100 Ry. The radius cutoff $r_c$ is chosen to
be 6 a.u.. As we can see, the radial functions of these atomic
orbitals have many oscillations. These oscillations are unphysical
and may lower the transferability of the atomic basis set. To get
rid of the alloying oscillations of the orbitals, at each step after
we obtain the orbitals which minimize the spillage, we add a
procedure to optimize the shape of the radial functions as follows.
We define the ``kinetic energy'' of an atomic orbital as,
\begin{equation}
T_{\mu}(c_{\mu q})=\sum_{q}c^2_{\mu q} q^2/2 + \kappa\, ,
\end{equation}
where $q$ satisfy $j_l(q r_c)$=0. $c_{\mu q}$ are the coefficients
of the spherical Bessel functions that normalize the atomic
orbitals, i.e., $\langle \phi_{\mu}|\phi_{\mu} \rangle$=1. $\kappa$
is a penalty function that
\begin{equation}
\kappa=\left\{
\begin{array}{ll}
0, &  \mathcal{S}/\mathcal{S}_0-1 < \Delta\\
\infty, &  \mathcal{S}/\mathcal{S}_0-1 > \Delta
\end{array}
\right. \, ,
\end{equation}
where $\mathcal{S}_{0}$ is the minimal spillage value for the given
orbital set, and $\mathcal{S}$ is the current spillage value for the
given coefficients $c_{\mu q}$ . We found $\Delta$=0.002 $\sim$ 0.005
can be sufficient to smooth out the atomic orbitals. The ``kinetic
energies'' of the atomic orbitals are also minimized via a simulated
annealing method. After the shape optimization, the final spillage
values will be slightly larger than the minimal ones, and
have little influence on the accuracy of these basis sets.

The shape optimized orbitals are plotted in Fig.\ref{fig:shape} (d),
(e), (f), for the $s$, $p$, $d$ orbitals respectively, compared to
those of unoptimized orbitals. As we see the alloying oscillations
in the original orbitals have been gotten rid of, and the shapes of the
optimized orbitals are much smoother than the origin ones, which
implies a better transferability. We calculate the total energies of
the reference dimmers using the optimized orbitals and find that
they offer the same accuracy for the selected dimers as the origin
ones.

\section{Results and discussion}
\label{sec:results}

In this section, we do intensive tests on the accuracy and
transferability of the atomic orbital basis sets generated using the
the scheme given in  Sec. \ref{sec:methods} for wide variety of
materials, including covalent, ionic, metallic systems. The lattice
constants, bulk modulus, band structures calculated from the atomic
basis are compared to those calculated from plane wave basis.
Especially, GaN, ZnO and Al are known to have several stable
structures that are energetically very close to each other, which
provide very good tests on the quality of the atomic basis sets.

All the calculations were
done using density functional methods\cite{hohenberg64,kohn65} (DFT)
with local density approximation(LDA) in Perdew-Zunger
form.\cite{perdew81} Norm conserving
pseudopotentials\cite{kleinman82} are used in fully separate
form.\cite{troullier91}
Periodic boundary condition is used for solids systems, and the
integration over Brillouin zone is replaced by sum up Monkhorst-Pack
$k$ points.\cite{monkhorst76}

\subsection{Cutoff radii of orbitals}

\begin{figure}
\centering
\includegraphics[width=3in]{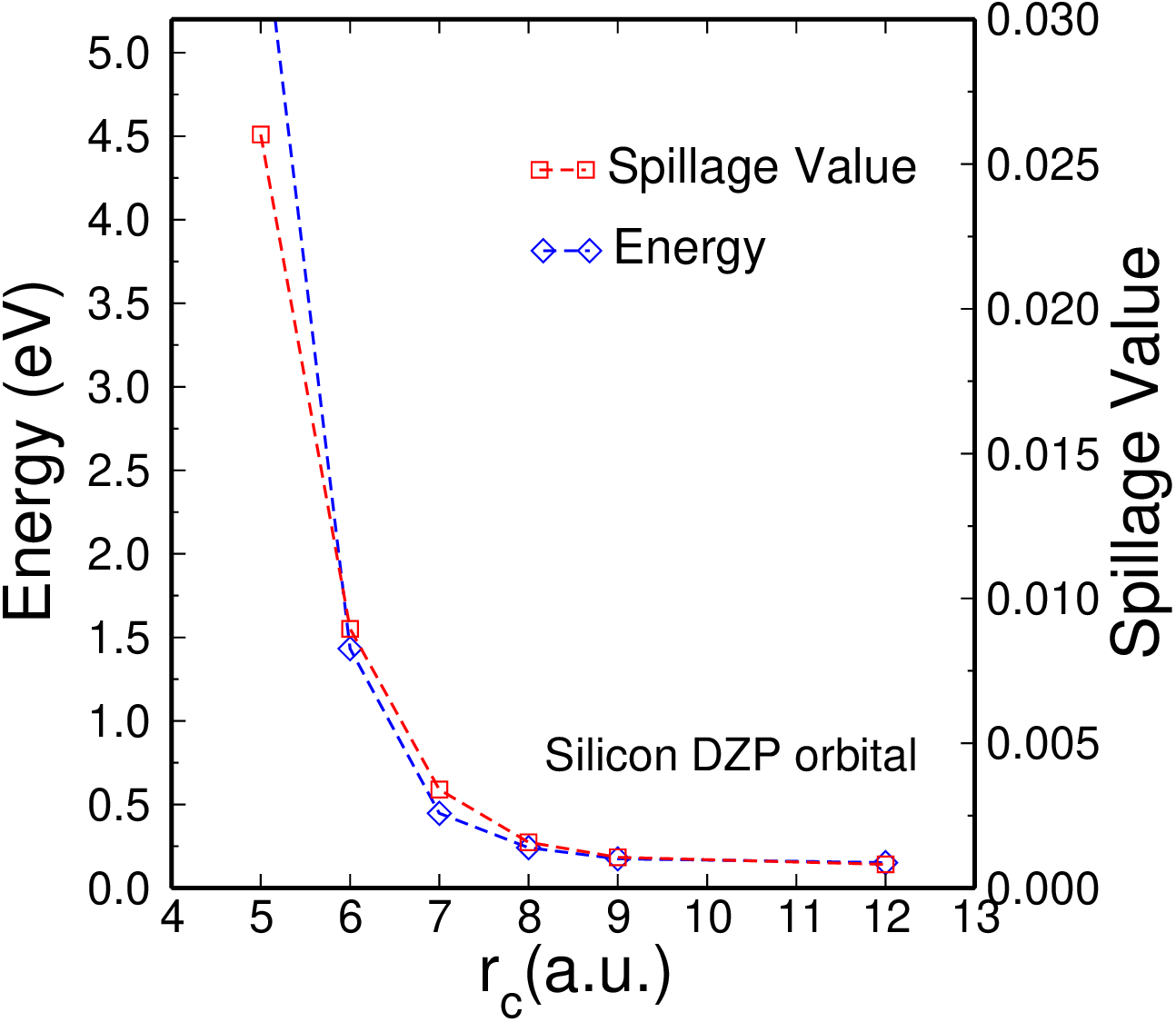}
\caption{(Color online)The total energy difference (blue line) and average
spillage value of five dimers (red line) as functions of orbital radius
cutoff $r_c$ for Si DZP orbitals. } \label{fig:Si_dzp_rcut}
\end{figure}

The range of radial function $r_c$ is one of the most important
parameters of the atomic orbitals. Usually larger $r_c$ leads to
more accurate results, but at the same time demands more
computational time and memory. One needs to choose proper $r_c$ that
balance the two factors. For 3D solid, the number of neighboring atoms
increases very fast as $r_c^3$, one has to use modest $r_c$, whereas
for 2D and 1D systems, $r_c$ can be relatively larger. It is also
important to balance the errors of different elements in the
system.\cite{anglada02}
As one shall see below, it is straightforward to use the spillage
value as a criterion to choose a proper $r_c$ for each
element.
Using the silicon diamond structure as an example. The energy cutoff
is chosen to be 50 Ry and the $k$ points are chosen to be
6$\times$6$\times$6, which are enough to converge the total
energies. In Fig.\ref{fig:Si_dzp_rcut} the blue curve is the total
energy difference between the Si DZP basis and the plane wave
calculations, whereas the red curve is the spillage, as a function
of $r_c$. As we see that both the spillage value and the total
energy decrease monotonically as $r_c$ increases and are almost on
top of each other. When $r_c$=6 a.u., the energy difference can be
reduced to about 1.5 eV per unit cell, and decreases rapidly to
about 0.25 eV per unit cell at  $r_c$=8 a.u.. The spillage values
for these two $r_c$ are about 9$\times$10$^{-3}$ and
2$\times$10$^{-3}$, respectively. Tests on other elements show
similar results, which clearly demonstrate that the spillage is an
excellent criterion for the quality of the atomic basis set.

\begin{figure}
\centering
\includegraphics[width=3in]{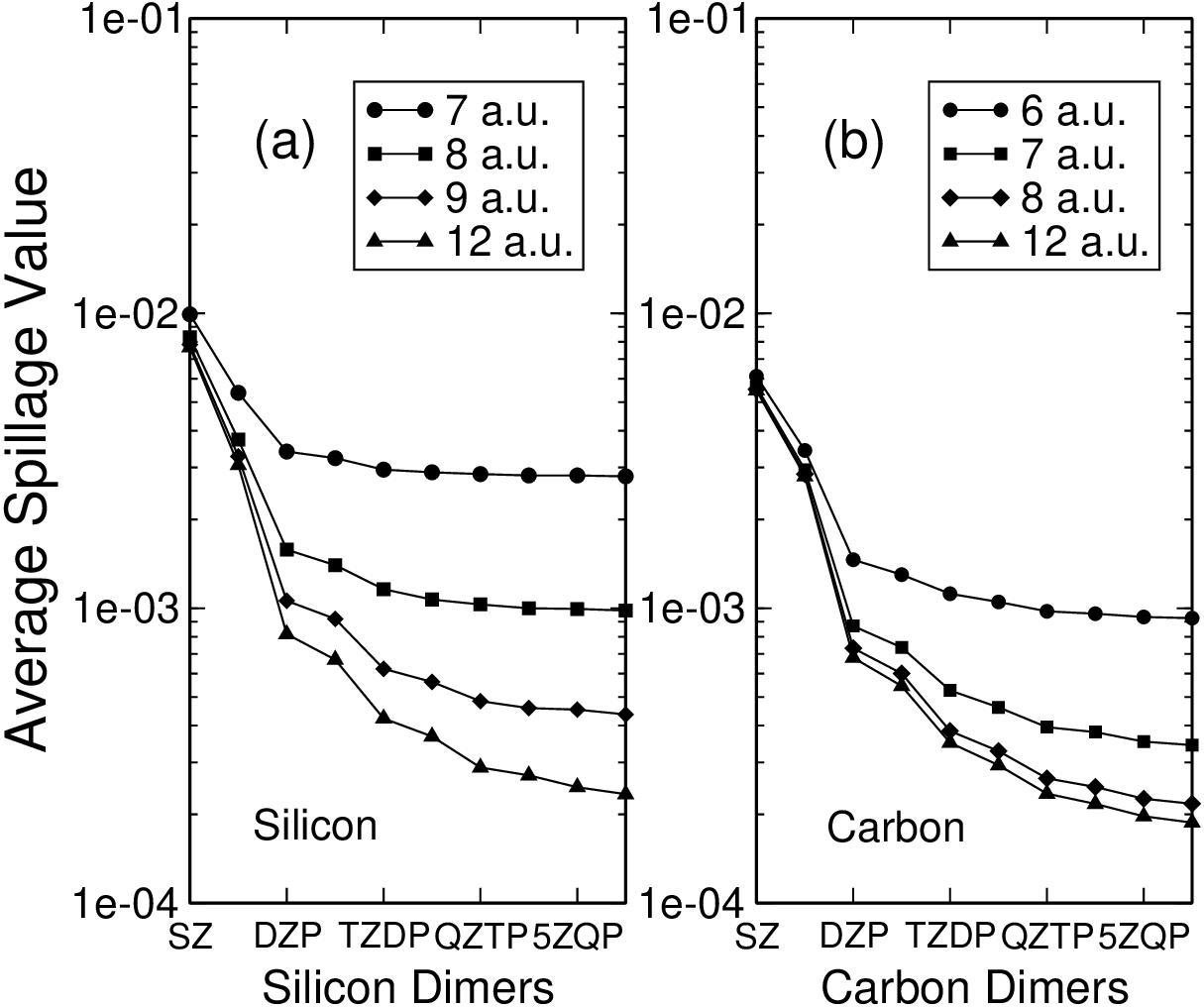}
\caption{Convergence of spillage value as functions of orbital radius
cutoff $r_c$ and basis size for (a) Si and (b)
Carbon dimers.} \label{fig:spillage}
\end{figure}

In order to further show how $r_c$ affects the spillage, we show in
Fig.\ref{fig:spillage} (a),(b) the spillage of different basis size
changes with $r_c$ for Si and C diamond structures. As we see, if
$r_c$ is too small, further increasing the size of the basis does
not significantly improve the quality of basis. For example, for Si
dimers, if one chose $r_c$=7 a.u., DZP basis has spillage about
3.5$\times$10$^{-3}$, further increasing the basis size does not
lower the spillage too much. However, increasing $r_c$ will
dramatically decrease the spillage. For $r_c$=12 a.u., the spillage
value of DZP basis can be as small as 8$\times$10$^{-4}$.
Figure~\ref{fig:spillage}(b) shows similar results for diamond.
However, as it is shown in the figure, carbon DZP orbitals with
$r_c$=6 a.u. are as good as DZP orbitals of silicon with $r_c$=8
a.u.. The lesson we can learn from the tests is that one should
choose proper basis set size for a given $r_c$.

\subsection{III-V and group IV semiconductors}

Semiconductors are a class of important materials. Here we test our
atomic bases for III-V and group IV semiconductors. The energy
cutoff is chosen to be 50 Ry and the $k$ points are chosen as
6$\times$6$\times$6, unless otherwise noticed. The results are
summarized in Table \ref{tab:semiconductor}. The $r_c$ of Ga, In,
Al, As, P, Sb, Ge elements are chosen to be 9 a.u. whereas $r_c$ of
silicon and carbon is chosen to be 8 a.u. and 6 a.u., respectively.
We use plane wave basis results as benchmark, also listed in
Table\ref{tab:semiconductor}. We can see that the results from the
single-zeta (SZ) basis (the minimal basis) have large deviations for
both lattice constants and bulk modulus. SZ basis predict too large
lattice constants than those calculated from plane wave basis for
more than 0.1 -0.2 a.u. and underestimate the bulk modulus for more
than $10\%$. However, modest size double-zeta plus polarized basis
(DZP) basis always offer good results. The largest difference occurs
in Ge, where the deviation of lattice constant and bulk modulus
between DZP and plane wave basis is 0.07 a.u. and 4 GPa,
respectively. After increasing the number of basis to triple-zeta
plus double polarized basis (TZDP, we also use notation 3s3p2d), we
can see systematically improvement over the DZP orbitals, and are
almost identical to those calculated from plane wave basis. The
atomic orbitals generated from dimer reference system can also
obtain such good results for solid systems, showing remarkable
transferability.

\begin{figure}
\centering
\includegraphics[width=3in]{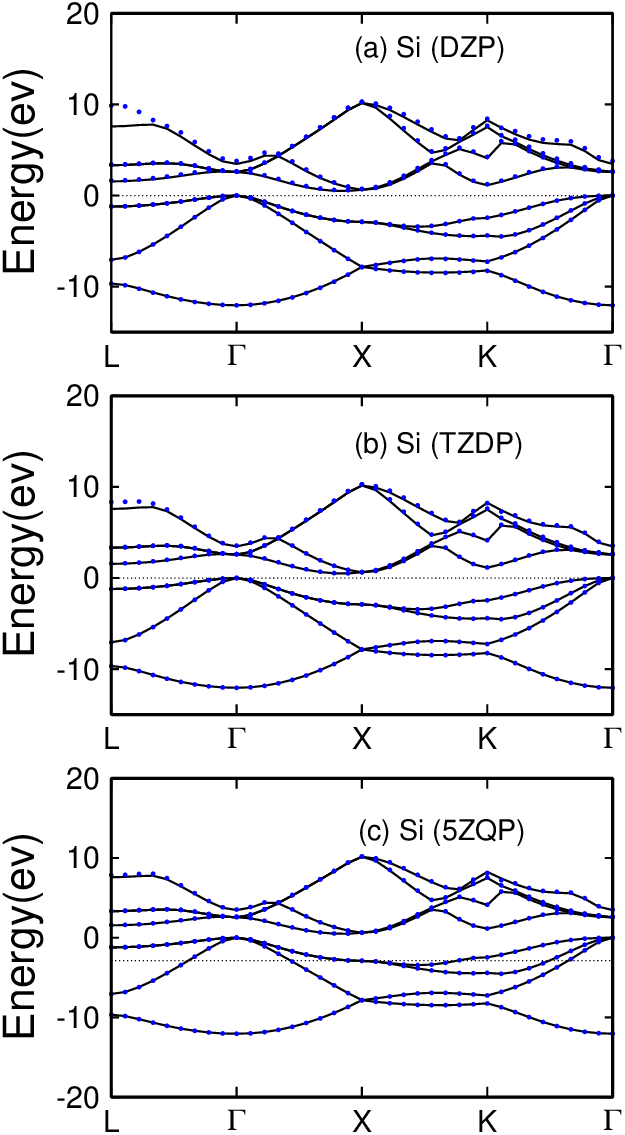}
\caption{(Color online) Comparision of the band structures of Si calculated by
different atomic bases (blue dotted lines) and plane wave basis (solid balck
lines).} \label{fig:bands}
\end{figure}

The band structures of silicon (diamond structure) are shown in
Fig.~\ref{fig:bands}. The lattice constant is fixed at 10.20 a.u.,
the energy cutoff is 50 Ry. We compare the band structures
calculated by atomic DZP, TZDP and 5ZQP (5s5p4d) bases in Fig.
\ref{fig:bands}(a)(b)(c) respectively and compared with the plane
wave results. The black solid lines represent the plane wave
results, whereas the blue dotted lines are the results of atomic
bases. The Fermi level is fixed at 0 eV. From Fig.\ref{fig:bands}(a)
we can see DZP basis already gives very good band structures, for
valence bands and the low energy conduction bands, except around $L$
point. The TZDP basis improves the band structures around $L$ point,
as illustrated in Fig.\ref{fig:bands}(b), though there is still
small difference. If we further increase the basis size to 5ZQP, we
can see the energy bands calculated from atomic orbitals are almost
identical to those from plane wave basis.

\begin{figure}
\centering
\includegraphics[width=3in]{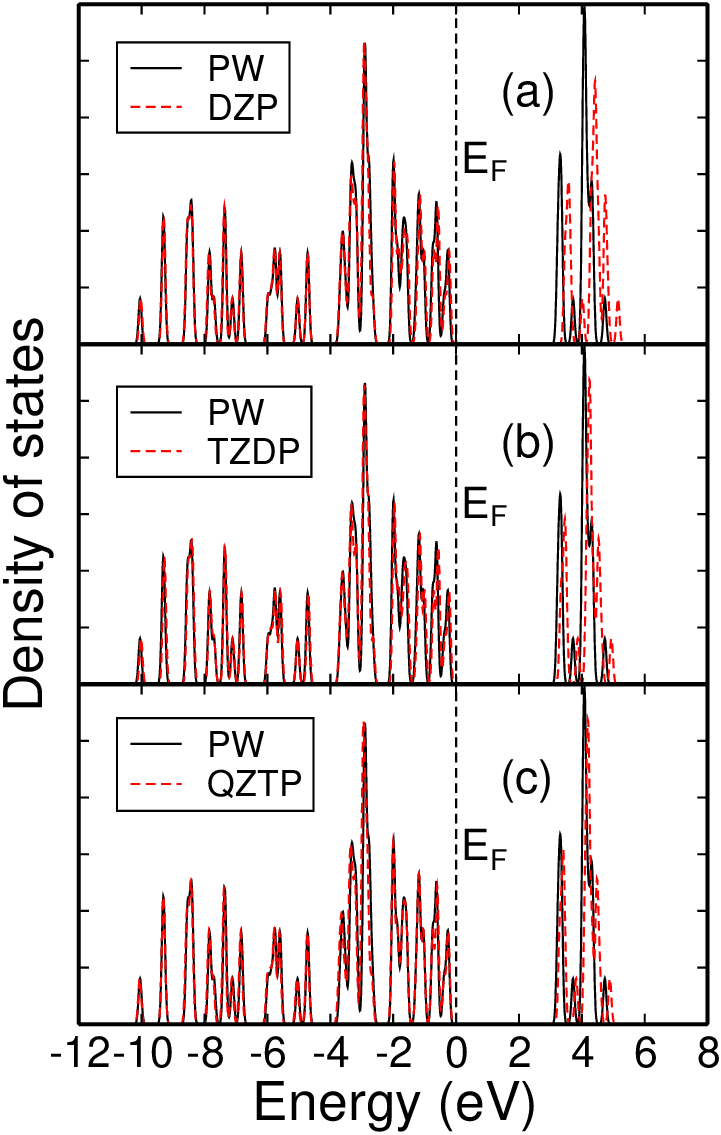}
\caption{(Color online)Comparison of the density of states (DOS) of
Si$_{29}$H$_{38}$ cluster calculated by different atomic bases and
plane wave basis.} \label{fig:Si29H38}
\end{figure}

We also calculate the electronic states of a Silicon cluster
containing 29 silicon atoms and 38 hydrogen atoms. We calculate
lowest 100 states and plot the density of states (DOS).
Also we denote 2s1p, 3s2p and 4s3p as
DZP, TZDP and QZTP basis for hydrogen, respectively. The results are shown in
Fig.~\ref{fig:Si29H38}(a), (b),(c) for the DZP, TZDP and QZTP bases
respectively. The DOS calculated by plane wave is shown in black
solid line, where as those calculated from atomic orbitals are shown
in red dashed lines. For the DZP basis, we find that the DOS of the
valence states are almost identical to that calculated from plane
wave basis. However, the DOS of conduction electrons shifts to the
higher energy side for about 280 meV
relative to the plane wave result. The TZDP basis improves the DZP's
results, however the DOS of conduction electrons still shifts a
little towards the higher energy side. QZTP further improves the
results, which are in excellent agreement with those calculated from
plane wave basis.

\subsection{GaN, ZnO, Al, MgO, Pb}

Now we test our atomic bases for several important materials,
including GaN, ZnO, Al and MgO and Pb. GaN, ZnO and Al have several
stable structures. Zn has 3d electrons as valence electrons, whereas
Al is metallic. They thus offer good examples for comprehensive
tests on the quality and transferability of the atomic bases.

\begin{figure}
\centering
\includegraphics[width=3in]{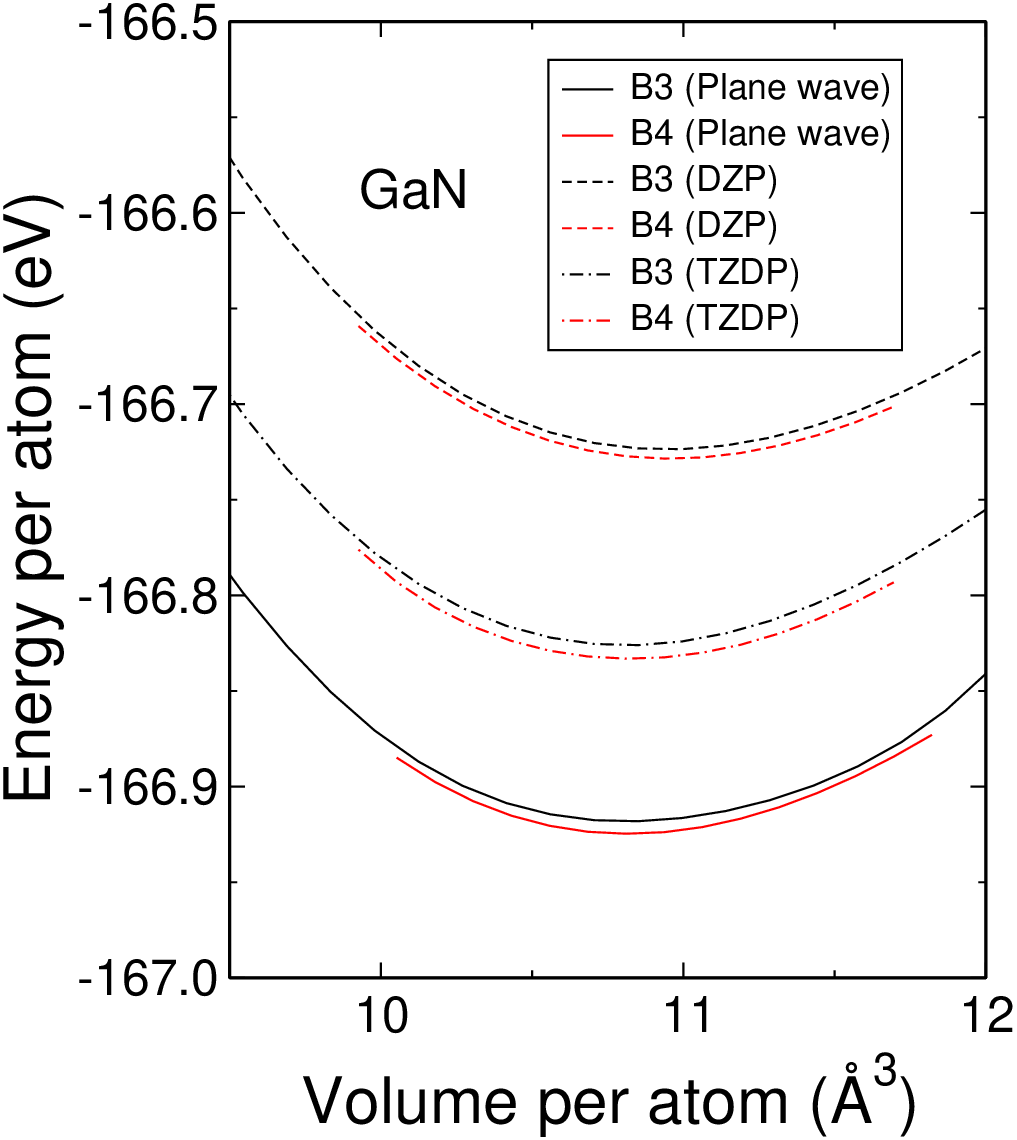}
\caption{(Color online) Comparison of the total energies of GaN zinc blende
structure (B3) and the wurtzite structure (B4) as functions of
volume per atom using different
  atomic bases to those using plane wave basis.} \label{fig:GaN_energy}
\end{figure}

GaN has two stable crystal structures, namely, the zinc blende
structure (B3) and the wurtzite structure (B4). The energy
difference between the two structures is very small. The B3
structure has only one structure parameter, i.e., the lattice
constant, whereas the B4 structure can be described by three
parameters: the lattice constants $a$, $c$ and the internal
parameter $u$, which describes the relative position of the two
hexagonal close-packed sublattices. The purpose of the test is to
see if the atomic basis set can predict correctly the energy
difference between the two structures. The energy cutoff is chosen
120 Ry and the $k$ points is chosen to be 6$\times$6$\times$6 for B3
and 6$\times$6$\times$4 for B4. The calculated total energies as
functions of volume per atom are shown in Fig.~\ref{fig:GaN_energy},
compared to those calculated from plane wave basis. The plane wave
calculations indicate that wurtzite structure is more stable than
zinc blende structure, which is also predicted correctly from both
DZP and TZDP basis sets. We can also see that the total energies are
systematically improved from DZP to TZDP. More properties including
structure parameters ($a$, $c$ and $u$), bulk modulus and the total
energies difference between the two structures are shown in
Table~\ref{tab:GaN}. We can see that DZP basis already give very
good structure parameters and bulk modulus compared with those
calculated from plane wave basis for both B3 and B4 structures. TZDP
basis further improves all properties. It gives much better bulk
modulus than DZP basis, which reduce the difference from about 10
GPa to less than 1 GPa. The total energy difference between zinc
blende structure and the wurtzite structure from plane wave
calculations is about 6 meV. Although the energy difference is very
small, both DZP and TZDP basis give rather good value, which are 4
meV and 7 meV, respectively. This proves that for GaN, local basis
can provide accurate results as good as plane wave basis.

\begin{figure}
\centering
\includegraphics[width=3in]{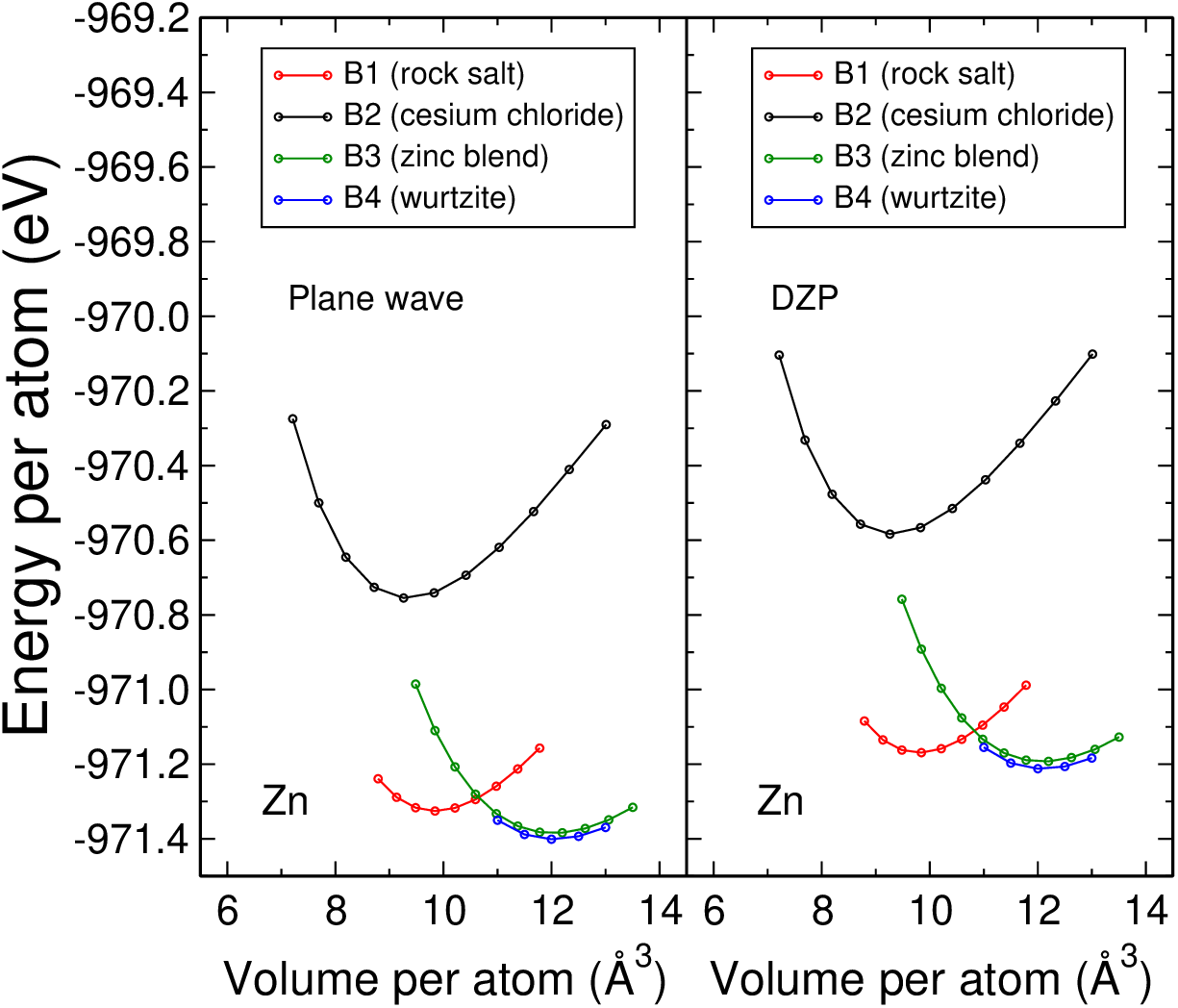}
\caption{(Color online) Comparison of the total energies of ZnO rock salt
structure (B1), cesium chloride structure (B2), zinc blende
structure (B3) and wurtzite structure (B4) as functions
volume per atom (a)
using plane wave basis to (b) those using atomic DZP basis.}
\label{fig:ZnO_energy}
\end{figure}

Let us see how the scheme work for systems containing 3d electrons.
Zinc is a transition-metal element, which contains 3d electrons. We
calculate four structures of ZnO, including rock salt structure
(B1), cesium chloride structure (B2), zinc blende structure (B3) and
wurtzite structure (B4).
The energy cutoff is chosen as 120 Ry,
and we use 6$\times$6$\times$6 $k$-meshes for B1, B2 and B3
structures and 6$\times$6$\times$4 $k$-meshes for B4 structure. We define 2s2p1d
for O and 2s1p2d for Zn as DZP basis, whereas 3s3p2d for O and
3s2p3d for Zn, as TZDP basis. The $r_{\rm cut}$ is 8 a.u. for O and
8 a.u. for Zn.
We plot total energies vs volume per atom for these four ZnO
structures in Fig. \ref{fig:ZnO_energy}(b) using DZP basis, compared
with the results calculated from plane wave basis shown in Fig.
\ref{fig:ZnO_energy}(a). Both bases give correct energy order for
the four structures compared to experiments. \cite{uddin06}
As we see the energy diagrams
calculated from DZP basis look almost the same as those calculated
from plane wave basis, except that all the total energies calculated
from DZP basis shift up for about 0.185 eV per atom relative to the
plane wave results. The calculations shows that the structures with
decreasing energy order is B2, B1, B3 and B4, and ground state
structure of ZnO is the wurtzite structure.

Table \ref{tab:ZnO} further shows the calculated structure
parameters, bulk modulus and energy differences of ZnO using DZP and
TZDP basis comparing with plane wave results. Both DZP and TZDP
basis give accurate lattice constants. The largest difference is
less than 0.02 \AA\, compared to plane wave results. However, TZDP
basis gives more accurate bulk modulus for all structures. The
energy difference between wurtzite phase and zinc blende phase is
also calculated accurately using different basis. The energy of
structure B4 is 9 meV per atom lower than that of the structure B3
as calculated by plane wave basis. DZP and TZDP give exact the same
results. We also show the lattice constant and bulk
modulus calculated by Gaussian basis. \cite{uddin06} We see that the
errors due to pseudopotentials and other approximations are bigger
than the errors caused by the atomic bases.

\begin{figure}
\centering
\includegraphics[width=3in]{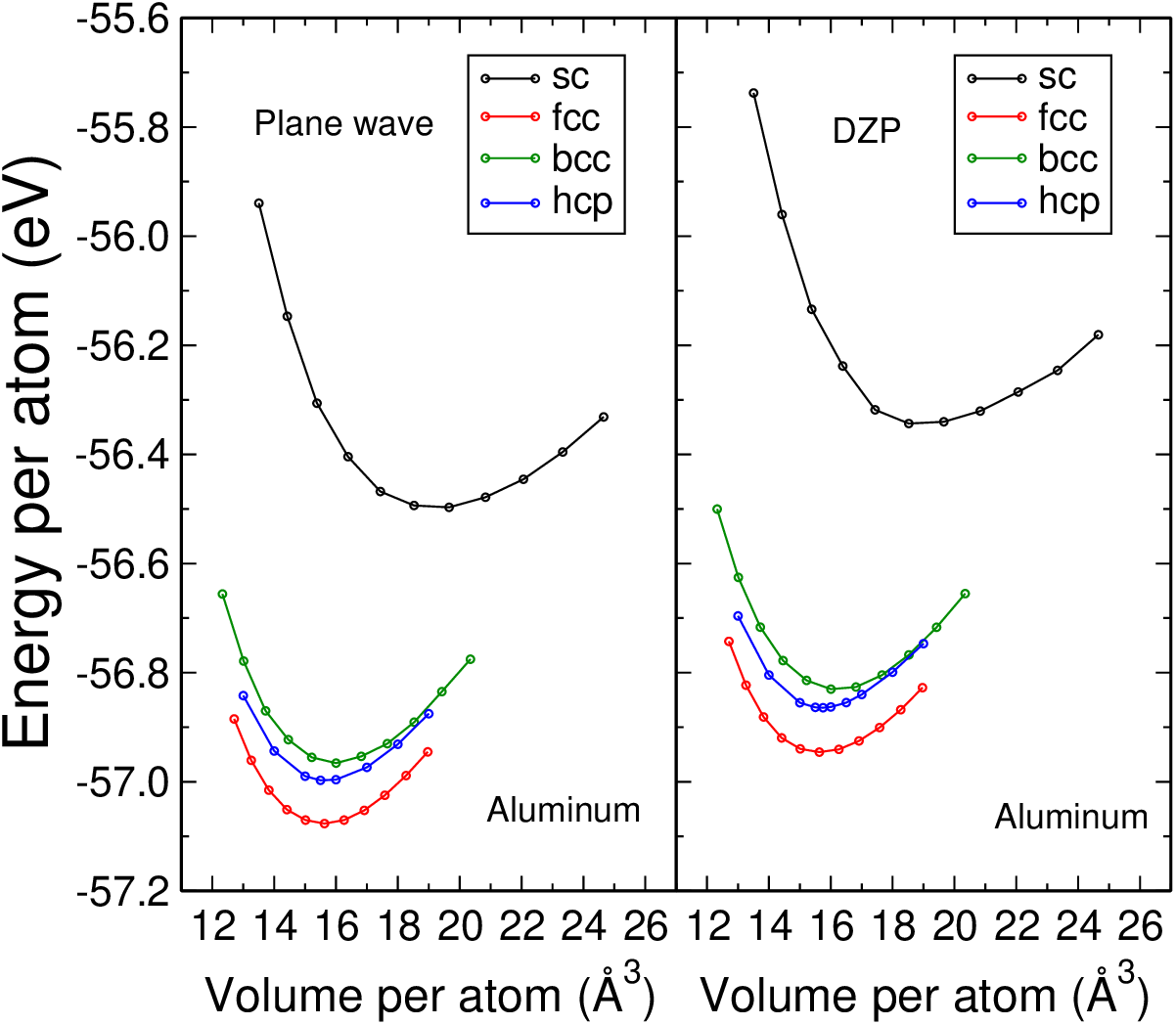}
\caption{(Color online) Comparison of the total energies of Al sc, fcc, bcc, hcp
structures as functions of volume per atom (a) using plane wave basis to
(b) those using atomic DZP basis.} \label{fig:Al_energy}
\end{figure}

Aluminum is a metal system. It is not obvious that atom-centered
atomic basis can describe it well. We test four structures of
Aluminum, including the simple cubic (sc) structure, the
face-centered cubic (fcc) structure, the body-centered cubic (bcc)
structure and the hexagonal-closed packed (hcp) structure. The
energy cutoff is fixed at 70 Ry, and the $k$ points are chosen as
6$\times$6$\times$6. Gaussian smearing is used in all calculations.
We use DZP (2s2p1d) orbitals for Aluminum with $r_c$=9 a.u. Figure
\ref{fig:Al_energy}(a), (b) compare the energy diagrams of the
Aluminum four structures calculated using plane wave basis and DZP
basis. As we see, the atomic basis provide excellent agreement with
the plane wave result for all structures. The fcc structure is the
lowest-energy structure of Aluminum predicted by both bases. We
summarize calculated properties, including lattice constant, bulk
modulus and energy difference between different structures of
Aluminum in table \ref{tab:Al}. Surprisingly, we find that DZP basis
can provide extremely good results compared to plane wave
calculations.\cite{kenny00}. For example, in our calculation, the
largest difference of calculated lattice constant is 0.011 \AA\, in
bcc structure, whereas the largest difference of bulk modulus is 1
GPa. The DZP basis can also give excellent energy differences
between different structures of Aluminum. These results are much
better than previous calculations, also using DZP basis.
\cite{kenny00} For example, in Ref. \onlinecite{kenny00}, the
lattice constants errors for fcc, hcp(a/c) bcc and sc are 0.044
\AA\, 0.054/0.040 \AA\, 0.037 \AA\, 0.013 \AA\, respectively. The
bulk modulus errors for fcc, hcp, bcc and sc is 4.9 GPa, 8.5GPa. 5.2
GPa and 2.9 GPa, respectively, which are much larger than those
obtained using our bases.

We also test the quality of our atomic bases for some other
materials, such as MgO, Pb, etc. The results are summarized in Table
\ref{tab:other}. The energy cutoff is 70 Ry for MgO and 50 Ry for
Pb. The $k$ points are 6$\times$6$\times$6. We all use 2s2p1d for
Mg, O, Pb. We use $r_c$=8 a.u. for Mg and O, and $r_c$=9 a.u. for Pb.
We compare the results with previous calculations using DZP
basis. \cite{junquera01} We find our basis give much
better bulk modulus for Pb than previous calculations.

\section{Summary}
\label{sec:conclusion}

We propose a unique scheme to construct fully optimized atomic basis
sets for density-functional calculations. The shape of the radial
functions are optimized by minimizing the {\it spillage} of the wave
functions between the atomic orbital calculations and the converged
plane wave calculations for dimer systems. Our method improves upon
that of previous methods in three aspects: (i) The shape of atomic
orbitals can be generated automatically without any pre-assumptions.
(ii) The atomic basis can be systematically improved within the same
framework. (iii)The transferability of atomic orbitals bases are
improved by carefully choosing the reference systems. The scheme is
easy to implement and very flexible. We have done extensive tests of
this scheme for wide variety of systems, including semiconductors,
ionic, covalent and metallic materials. The results show that the
obtained atomic basis sets are very satisfactory for both accuracy
and transferability.

\acknowledgments
L.H. acknowledges the support from the Chinese National Fundamental
Research Program 2006CB921900, the Innovation funds and ``Hundreds
of Talents'' program from Chinese Academy of Sciences.


\bibliographystyle{prsty}
\bibliographystyle{apsrev}
\bibliography{DFT}

\begin{thebibliography}{27}
\expandafter\ifx\csname natexlab\endcsname\relax\def\natexlab#1{#1}\fi
\expandafter\ifx\csname bibnamefont\endcsname\relax
  \def\bibnamefont#1{#1}\fi
\expandafter\ifx\csname bibfnamefont\endcsname\relax
  \def\bibfnamefont#1{#1}\fi
\expandafter\ifx\csname citenamefont\endcsname\relax
  \def\citenamefont#1{#1}\fi
\expandafter\ifx\csname url\endcsname\relax
  \def\url#1{\texttt{#1}}\fi
\expandafter\ifx\csname urlprefix\endcsname\relax\def\urlprefix{URL }\fi
\providecommand{\bibinfo}[2]{#2}
\providecommand{\eprint}[2][]{\url{#2}}

\bibitem[{\citenamefont{Otsuka et~al.}(2008)\citenamefont{Otsuka, Miyazaki,
  Ohno, Bowler, and Gillan}}]{otsuka08}
\bibinfo{author}{\bibfnamefont{T.}~\bibnamefont{Otsuka}},
  \bibinfo{author}{\bibfnamefont{T.}~\bibnamefont{Miyazaki}},
  \bibinfo{author}{\bibfnamefont{T.}~\bibnamefont{Ohno}},
  \bibinfo{author}{\bibfnamefont{D.~R.} \bibnamefont{Bowler}},
  \bibnamefont{and} \bibinfo{author}{\bibfnamefont{M.~J.}
  \bibnamefont{Gillan}}, \bibinfo{journal}{J. Phys.:Condens. Matter}
  \textbf{\bibinfo{volume}{20}}, \bibinfo{pages}{294201}
  (\bibinfo{year}{2008}).

\bibitem[{\citenamefont{Blum et~al.}(2009)\citenamefont{Blum, Gehrke, Hanke,
  Havu, Havu, Ren, Reuter, and Scheffler}}]{volker08}
\bibinfo{author}{\bibfnamefont{V.}~\bibnamefont{Blum}},
  \bibinfo{author}{\bibfnamefont{R.}~\bibnamefont{Gehrke}},
  \bibinfo{author}{\bibfnamefont{F.}~\bibnamefont{Hanke}},
  \bibinfo{author}{\bibfnamefont{P.}~\bibnamefont{Havu}},
  \bibinfo{author}{\bibfnamefont{V.}~\bibnamefont{Havu}},
  \bibinfo{author}{\bibfnamefont{X.}~\bibnamefont{Ren}},
  \bibinfo{author}{\bibfnamefont{K.}~\bibnamefont{Reuter}}, \bibnamefont{and}
  \bibinfo{author}{\bibfnamefont{M.}~\bibnamefont{Scheffler}},
  \bibinfo{journal}{Com. \ Phy. \ Commu.} \textbf{\bibinfo{volume}{06}},
  \bibinfo{pages}{022} (\bibinfo{year}{2009}).

\bibitem[{\citenamefont{Skylaris et~al.}(2008)\citenamefont{Skylaris, Haynes,
  Mostofi, and Payne}}]{skylaris08}
\bibinfo{author}{\bibfnamefont{C.-K.} \bibnamefont{Skylaris}},
  \bibinfo{author}{\bibfnamefont{P.~D.} \bibnamefont{Haynes}},
  \bibinfo{author}{\bibfnamefont{A.~A.} \bibnamefont{Mostofi}},
  \bibnamefont{and} \bibinfo{author}{\bibfnamefont{M.~C.} \bibnamefont{Payne}},
  \bibinfo{journal}{J. Phys.:Condens. Matter} \textbf{\bibinfo{volume}{20}},
  \bibinfo{pages}{064209} (\bibinfo{year}{2008}).

\bibitem[{\citenamefont{Artacho et~al.}(2008)\citenamefont{Artacho, Anglada,
  Dieguez, Gale, Garcia, Junquera, Martin, Ordejon, Pruneda, Sanchez-Portal
  et~al.}}]{artacho08}
\bibinfo{author}{\bibfnamefont{E.}~\bibnamefont{Artacho}},
  \bibinfo{author}{\bibfnamefont{E.}~\bibnamefont{Anglada}},
  \bibinfo{author}{\bibfnamefont{O.}~\bibnamefont{Dieguez}},
  \bibinfo{author}{\bibfnamefont{J.~D.} \bibnamefont{Gale}},
  \bibinfo{author}{\bibfnamefont{A.}~\bibnamefont{Garcia}},
  \bibinfo{author}{\bibfnamefont{J.}~\bibnamefont{Junquera}},
  \bibinfo{author}{\bibfnamefont{R.~M.} \bibnamefont{Martin}},
  \bibinfo{author}{\bibfnamefont{P.}~\bibnamefont{Ordejon}},
  \bibinfo{author}{\bibfnamefont{J.~M.} \bibnamefont{Pruneda}},
  \bibinfo{author}{\bibfnamefont{D.}~\bibnamefont{Sanchez-Portal}},
  \bibnamefont{et~al.}, \bibinfo{journal}{J. Phys.:Condens. Matter}
  \textbf{\bibinfo{volume}{20}}, \bibinfo{pages}{064209}
  (\bibinfo{year}{2008}).

\bibitem[{\citenamefont{Goedecker}(1999)}]{goedecker99}
\bibinfo{author}{\bibfnamefont{S.}~\bibnamefont{Goedecker}},
  \bibinfo{journal}{Rev. \ Mod. \ Phys.} \textbf{\bibinfo{volume}{71}},
  \bibinfo{pages}{1085} (\bibinfo{year}{1999}).

\bibitem[{\citenamefont{Kohn}(1996)}]{kohn96}
\bibinfo{author}{\bibfnamefont{W.}~\bibnamefont{Kohn}}, \bibinfo{journal}{Phys.
  \ Rev. \ Lett} \textbf{\bibinfo{volume}{76}}, \bibinfo{pages}{3168}
  (\bibinfo{year}{1996}).

\bibitem[{\citenamefont{Soler et~al.}(2002)\citenamefont{Soler, Artacho, Gale,
  Garcia, Junquera, Ordejon, and Portal}}]{soler02}
\bibinfo{author}{\bibfnamefont{J.~M.} \bibnamefont{Soler}},
  \bibinfo{author}{\bibfnamefont{E.}~\bibnamefont{Artacho}},
  \bibinfo{author}{\bibfnamefont{J.~D.} \bibnamefont{Gale}},
  \bibinfo{author}{\bibfnamefont{A.}~\bibnamefont{Garcia}},
  \bibinfo{author}{\bibfnamefont{J.}~\bibnamefont{Junquera}},
  \bibinfo{author}{\bibfnamefont{P.}~\bibnamefont{Ordejon}}, \bibnamefont{and}
  \bibinfo{author}{\bibfnamefont{D.~S.} \bibnamefont{Portal}},
  \bibinfo{journal}{J. Phys:: Condenss. Matter} \textbf{\bibinfo{volume}{14}},
  \bibinfo{pages}{2745} (\bibinfo{year}{2002}).

\bibitem[{\citenamefont{Ozaki}(2003)}]{ozaki03}
\bibinfo{author}{\bibfnamefont{T.}~\bibnamefont{Ozaki}},
  \bibinfo{journal}{Phys. \ Rev. \ B} \textbf{\bibinfo{volume}{67}},
  \bibinfo{pages}{155108} (\bibinfo{year}{2003}).

\bibitem[{\citenamefont{Sankey and Niklewski}(1989)}]{sankey89}
\bibinfo{author}{\bibfnamefont{O.~F.} \bibnamefont{Sankey}} \bibnamefont{and}
  \bibinfo{author}{\bibfnamefont{D.~J.} \bibnamefont{Niklewski}},
  \bibinfo{journal}{Phys. \ Rev. \ B} \textbf{\bibinfo{volume}{40}},
  \bibinfo{pages}{3979} (\bibinfo{year}{1989}).

\bibitem[{\citenamefont{Junquera et~al.}(2001)\citenamefont{Junquera, Paz,
  Sanchez-Portal, and Artacho}}]{junquera01}
\bibinfo{author}{\bibfnamefont{J.}~\bibnamefont{Junquera}},
  \bibinfo{author}{\bibfnamefont{O.}~\bibnamefont{Paz}},
  \bibinfo{author}{\bibfnamefont{D.}~\bibnamefont{Sanchez-Portal}},
  \bibnamefont{and} \bibinfo{author}{\bibfnamefont{E.}~\bibnamefont{Artacho}},
  \bibinfo{journal}{Phys. \ Rev. \ B} \textbf{\bibinfo{volume}{64}},
  \bibinfo{pages}{235111} (\bibinfo{year}{2001}).

\bibitem[{\citenamefont{Portal et~al.}(1995)\citenamefont{Portal, Artacho, and
  Soler}}]{portal95}
\bibinfo{author}{\bibfnamefont{D.~S.} \bibnamefont{Portal}},
  \bibinfo{author}{\bibfnamefont{E.}~\bibnamefont{Artacho}}, \bibnamefont{and}
  \bibinfo{author}{\bibfnamefont{J.~M.} \bibnamefont{Soler}},
  \bibinfo{journal}{Solid State Communications} \textbf{\bibinfo{volume}{95}},
  \bibinfo{pages}{685} (\bibinfo{year}{1995}).

\bibitem[{\citenamefont{Portal et~al.}(1996)\citenamefont{Portal, Artacho, and
  Soler}}]{portal96}
\bibinfo{author}{\bibfnamefont{D.~S.} \bibnamefont{Portal}},
  \bibinfo{author}{\bibfnamefont{E.}~\bibnamefont{Artacho}}, \bibnamefont{and}
  \bibinfo{author}{\bibfnamefont{J.~M.} \bibnamefont{Soler}},
  \bibinfo{journal}{J. Phys: Condes Matter} \textbf{\bibinfo{volume}{8}},
  \bibinfo{pages}{3859} (\bibinfo{year}{1996}).

\bibitem[{\citenamefont{Kenny et~al.}(2000)\citenamefont{Kenny, Horsfield, and
  Fujitani}}]{kenny00}
\bibinfo{author}{\bibfnamefont{S.~D.} \bibnamefont{Kenny}},
  \bibinfo{author}{\bibfnamefont{A.~P.} \bibnamefont{Horsfield}},
  \bibnamefont{and} \bibinfo{author}{\bibfnamefont{H.}~\bibnamefont{Fujitani}},
  \bibinfo{journal}{Phys. \ Rev. \ B} \textbf{\bibinfo{volume}{62}},
  \bibinfo{pages}{4899} (\bibinfo{year}{2000}).

\bibitem[{\citenamefont{Gusso}(2008)}]{gusso08}
\bibinfo{author}{\bibfnamefont{M.}~\bibnamefont{Gusso}}, \bibinfo{journal}{J.
  Chem. Phys.} \textbf{\bibinfo{volume}{128}}, \bibinfo{pages}{044102(R)}
  (\bibinfo{year}{2008}).

\bibitem[{\citenamefont{Hohenberg and Kohn}(1964)}]{hohenberg64}
\bibinfo{author}{\bibfnamefont{P.}~\bibnamefont{Hohenberg}} \bibnamefont{and}
  \bibinfo{author}{\bibfnamefont{W.}~\bibnamefont{Kohn}},
  \bibinfo{journal}{Phys. \ Rev.} \textbf{\bibinfo{volume}{136}},
  \bibinfo{pages}{864B} (\bibinfo{year}{1964}).

\bibitem[{\citenamefont{Kohn and Sham}(1965)}]{kohn65}
\bibinfo{author}{\bibfnamefont{W.}~\bibnamefont{Kohn}} \bibnamefont{and}
  \bibinfo{author}{\bibfnamefont{L.~J.} \bibnamefont{Sham}},
  \bibinfo{journal}{Phys. \ Rev.} \textbf{\bibinfo{volume}{140}},
  \bibinfo{pages}{1133A} (\bibinfo{year}{1965}).

\bibitem[{\citenamefont{Perdew and Zunger}(1981)}]{perdew81}
\bibinfo{author}{\bibfnamefont{J.~P.} \bibnamefont{Perdew}} \bibnamefont{and}
  \bibinfo{author}{\bibfnamefont{A.}~\bibnamefont{Zunger}},
  \bibinfo{journal}{Phys. \ Rev. \ B} \textbf{\bibinfo{volume}{23}},
  \bibinfo{pages}{5048} (\bibinfo{year}{1981}).

\bibitem[{\citenamefont{Kleinman and Bylander}(1982)}]{kleinman82}
\bibinfo{author}{\bibfnamefont{L.}~\bibnamefont{Kleinman}} \bibnamefont{and}
  \bibinfo{author}{\bibfnamefont{D.~M.} \bibnamefont{Bylander}},
  \bibinfo{journal}{Phys. \ Rev. \ Lett} \textbf{\bibinfo{volume}{48}},
  \bibinfo{pages}{1425} (\bibinfo{year}{1982}).

\bibitem[{\citenamefont{Troullier and Martins}(1991)}]{troullier91}
\bibinfo{author}{\bibfnamefont{N.}~\bibnamefont{Troullier}} \bibnamefont{and}
  \bibinfo{author}{\bibfnamefont{J.~L.} \bibnamefont{Martins}},
  \bibinfo{journal}{Phys. \ Rev. \ B} \textbf{\bibinfo{volume}{43}},
  \bibinfo{pages}{1993} (\bibinfo{year}{1991}).

\bibitem[{\citenamefont{Monkhorst and Pack}(1976)}]{monkhorst76}
\bibinfo{author}{\bibfnamefont{H.~J.} \bibnamefont{Monkhorst}}
  \bibnamefont{and} \bibinfo{author}{\bibfnamefont{J.~D.} \bibnamefont{Pack}},
  \bibinfo{journal}{Phys. \ Rev. \ B} \textbf{\bibinfo{volume}{13}},
  \bibinfo{pages}{5188} (\bibinfo{year}{1976}).

\bibitem[{\citenamefont{Anglada et~al.}(2002)\citenamefont{Anglada, M.Soler,
  Junquera, and Artacho}}]{anglada02}
\bibinfo{author}{\bibfnamefont{E.}~\bibnamefont{Anglada}},
  \bibinfo{author}{\bibfnamefont{J.}~\bibnamefont{M.Soler}},
  \bibinfo{author}{\bibfnamefont{J.}~\bibnamefont{Junquera}}, \bibnamefont{and}
  \bibinfo{author}{\bibfnamefont{E.}~\bibnamefont{Artacho}},
  \bibinfo{journal}{Phys. \ Rev. \ B} \textbf{\bibinfo{volume}{66}},
  \bibinfo{pages}{205101} (\bibinfo{year}{2002}).

\bibitem[{\citenamefont{Uddin and Scuseria}(2006)}]{uddin06}
\bibinfo{author}{\bibfnamefont{J.}~\bibnamefont{Uddin}} \bibnamefont{and}
  \bibinfo{author}{\bibfnamefont{G.~E.} \bibnamefont{Scuseria}},
  \bibinfo{journal}{Phys. \ Rev. \ B} \textbf{\bibinfo{volume}{74}},
  \bibinfo{pages}{245115} (\bibinfo{year}{2006}).

\bibitem[{\citenamefont{Vurgaftman et~al.}(2001)\citenamefont{Vurgaftman,
  Meyer, and Ram-Mohan}}]{vurgaftman01}
\bibinfo{author}{\bibfnamefont{I.}~\bibnamefont{Vurgaftman}},
  \bibinfo{author}{\bibfnamefont{J.~R.} \bibnamefont{Meyer}}, \bibnamefont{and}
  \bibinfo{author}{\bibfnamefont{L.~R.} \bibnamefont{Ram-Mohan}},
  \bibinfo{journal}{Jour. \ App. \ Phys} \textbf{\bibinfo{volume}{89}},
  \bibinfo{pages}{5815} (\bibinfo{year}{2001}).

\bibitem[{\citenamefont{C.Kittle}(1986)}]{kittle86}
\bibinfo{author}{\bibnamefont{C.Kittle}}, \bibinfo{journal}{Introduction to
  Solid State Physics}  (\bibinfo{year}{1986}).

\bibitem[{\citenamefont{Juan and Kaxiras}(1993)}]{juan93}
\bibinfo{author}{\bibfnamefont{Y.-M.} \bibnamefont{Juan}} \bibnamefont{and}
  \bibinfo{author}{\bibfnamefont{E.}~\bibnamefont{Kaxiras}},
  \bibinfo{journal}{Phys. \ Rev. \ B} \textbf{\bibinfo{volume}{48}},
  \bibinfo{pages}{14944} (\bibinfo{year}{1993}).

\bibitem[{\citenamefont{Cohen}(1985)}]{cohen85}
\bibinfo{author}{\bibfnamefont{M.~L.} \bibnamefont{Cohen}},
  \bibinfo{journal}{Phys. \ Rev. \ B} \textbf{\bibinfo{volume}{32}},
  \bibinfo{pages}{7988} (\bibinfo{year}{1985}).

\bibitem[{\citenamefont{Karch et~al.}(1998)\citenamefont{Karch, Wagner, and
  Bechstedt}}]{karch98}
\bibinfo{author}{\bibfnamefont{K.}~\bibnamefont{Karch}},
  \bibinfo{author}{\bibfnamefont{J.~M.} \bibnamefont{Wagner}},
  \bibnamefont{and}
  \bibinfo{author}{\bibfnamefont{F.}~\bibnamefont{Bechstedt}},
  \bibinfo{journal}{Phys. \ Rev. \ B} \textbf{\bibinfo{volume}{57}},
  \bibinfo{pages}{7043} (\bibinfo{year}{1998}).

\end{thebibliography}

\clearpage
\newpage
\begin{table}
\caption{Comparison of the calculated lattice constants $a$ (in a.u.) and bulk
modulus $B$ (in GPa) using plane wave basis (PW) and atomic bases for typical III-V and
group IV materials.}\label{tab:semiconductor}
\begin{tabular}{ccccccccccc}
\hline \hline
 & \mc{5}{c} {$a$ } & \mc{5}{c}{$B$} \\
Compound & SZ & DZP & TZDP & PW & Expr. & SZ & DZP & TZDP & PW & Expr. \\
\hline
GaAs & 10.67 & 10.50 & 10.49& 10.48 & 10.68\footnotemark[1] & 69 & 78 & 77& 77 & 75.57\footnotemark[3]\\
GaP  & 10.28 & 10.11 & 10.11& 10.10 & 10.30\footnotemark[1] & 82 & 92 & 93& 93 & 89\footnotemark[4]\\
GaSb & 11.54 & 11.38 & 11.37& 11.36 & 11.52\footnotemark[1] & 49 & 59 & 58& 57 & 57\footnotemark[4]\\
InAs & 11.40 & 11.27 & 11.28& 11.28 & 11.45\footnotemark[1] & 63 & 66 & 65& 65 & 60\footnotemark[4]\\
InP  & 11.07 & 10.94 & 10.94& 10.93 & 11.09\footnotemark[1] & 78 & 79 & 79& 80 & 71 \footnotemark[4]\\
InSb & 12.33 & 12.05 & 12.05& 12.07 & 12.24\footnotemark[1] & 41 & 50 & 49& 50 & 47\footnotemark[4]\\
AlAs & 10.88 & 10.63 & 10.62& 10.59 & 10.70\footnotemark[1] & 67 & 76 & 76& 76 & 77\footnotemark[4]\\
AlP  & 10.50 & 10.26 & 10.25& 10.21 & 10.33\footnotemark[1] & 64 & 87 & 88& 89 & 86\footnotemark[4]\\
AlSb & 11.83 & 11.58 & 11.57& 11.54 & 11.59\footnotemark[1] & 48 & 55 & 56& 56 & 58\footnotemark[4]\\
Ge   & 10.82 & 10.68 & 10.61& 10.61 & 10.70\footnotemark[3] & 57 & 67 & 71& 71 & 77.20\footnotemark[3]\\
Si   & 10.59 & 10.28 & 10.25& 10.23 & 10.26\footnotemark[2] & 74 & 94 & 94& 94 & 99\footnotemark[2]\\
C\footnotemark[5] & 6.78  & 6.67  & 6.67 & 6.67  & 6.75 \footnotemark[2] &436 & 470 & 467& 466& 442\footnotemark[2]\\
\hline  \hline
\footnotetext[1]{I. Vurgaftman, J. R. Meyer and L. R. Ram-Mohan, Ref. \onlinecite{vurgaftman01}.}
\footnotetext[2]{C. Kittel, Ref. \onlinecite{kittle86}.}
\footnotetext[3]{Yu-Min Juan and Efthimios Kaxiras, Ref.\onlinecite{juan93}.}
\footnotetext[4]{Marvin L. Cohen, Ref.\onlinecite{cohen85}.}
\footnotetext[5]{Energy cutoff 100 Ry is used.}
\end{tabular}
\end{table}
%
\clearpage
\newpage
\begin{table}
\caption{Basis comparision for GaN Zinc blende (B3) and wurtzite (B4)
structures. $a$, $c$ (in \AA) are the lattice constants, and $c$ (in \AA)
is the internal parameter. $B$ (in GPa) is the bulk modulus, whereas
$\Delta E$ (in eV/atom)  is the total energy difference between
different structures.
The energy difference is set to zero for B4 structure.
The data of ``Other calculations''
and experimental values are taken from the reference \onlinecite{karch98}.}
\label{tab:GaN}
\begin{tabular}{ccccccccc}
\hline \hline
Compound & Properties & PW   & DZP    & TZDP & Other calculations
 & Experiment\\
\hline
GaN(B3) & $a$         & 4.424  & 4.441  & 4.422  & 4.446$\sim$4.46 & 4.49\\
        & $B$         & 207    & 197    & 208    & 184$\sim$207 & 173\\
        & $\Delta$E & 0.006  & 0.004  & 0.007\\
GaN(B4) & $a$         & 3.130  & 3.142  & 3.130  & 3.126$\sim$3.170 & 3.160$\sim$3.189\\
        & $c/a$       & 1.629  & 1.631  & 1.630  & 1.620$\sim$1.638 & 1.621$\sim$1.626\\
        & $u$         & 0.377  & 0.377  & 0.377  & 0.377$\sim$0.379 & \\
        & $B$         & 207    & 194    & 207    & 190$\sim$245 & 188$\sim$237\\
        & $\Delta$E & 0 & 0 & 0\\
\hline  \hline
\end{tabular}
\end{table}
%
\clearpage
\newpage
\begin{table}
\caption{Basis comparison for ZnO
rock salt (B1), cesium chloride (B2), zinc blende
(B3) and wurtzite (B4) structures. $a$, $c$ (in \AA) are the lattice constants, and $c$ (in \AA)
is the internal parameter. $B$ (in GPa) is the bulk modulus, whereas
$\Delta E$ (in eV/atom) is the total energy difference between
different structures.
The energy difference is set to zero for B4 structure.
The results from Gaussian orbitals and the experimental values are taken form
Ref. \onlinecite{uddin06}. }\label{tab:ZnO}
\begin{tabular}{cccccccc}
\hline \hline
Compound   & Properties & PW     & DZP    & TZDP    & Gaussian
& Experiment\\
\hline
ZnO(B1)    & $a$          & 4.286  & 4.296  & 4.296   & 4.218 & 4.271-4.283 \\
           & $B$          & 196    & 198    & 195     & 203   & 177-228 \\
           & $\Delta E$ & 0.068  & 0.066  & 0.058   &       &  \\
ZnO(B2)    & $a$          & 2.653  & 2.659  & 2.658   & 2.614 &  \\
           & $B$          & 194    & 190    & 191     & 201   &  \\
           & $\Delta E$ & 0.639  & 0.647  & 0.620   &       &  \\
ZnO(B3)    & $a$          & 4.583  & 4.596  & 4.595   & 4.509 & 4.62  \\
           & $B$          & 152    & 150    & 151     & 154   &   \\
           & $\Delta E$ & 0.009  & 0.009  & 0.009   &       &   \\
ZnO(B4)    & $a$          & 3.256  & 3.266  & 3.264   & 3.205 & 3.248-3.250\\
           & $c$          & 5.246  & 5.256  & 5.255   & 5.151 & 5.207-5.210\\
           & $u$          & 0.381  & 0.381  & 0.381   & 0.381 & 0.382 \\
           & $B$          & 152    & 150    & 151     & 155   & 136-183\\
           & $\Delta E$ & 0  & 0  & 0   &       & \\
\hline  \hline
\end{tabular}
\end{table}
%
\clearpage
\newpage
\begin{table}
\caption{
Comparison of the calculated lattice constants $a$, $c$ (in \AA),
bulk modulus $B$ (in GPa) and energy difference $\Delta E$ (in eV/atom)
for Aluminum fcc, bcc, sc and hcp strcutures, using different basis. The
energy difference is set to zero for fcc structure.
The results of ``Other DZP'' and
``Other PW'' are taken from Ref. \onlinecite{kenny00}.}\label{tab:Al}
\begin{tabular}{cccccc}
\hline \hline
Compound & Properties & PW & DZP & Other DZP  & Other PW \\
\hline
Al(fcc) & $a$ & 3.964 & 3.974 & 4.011 & 3.967\\
& $B$ & 80  & 79 & 76.2 & 81.1\\
& $\Delta E$ & 0 & 0  & 0  & 0\\
Al(bcc) & $a$ & 3.175 & 3.186 & 3.212 & 3.175\\
& $B$ & 75 & 75 & 67.1 & 72.3\\
& $\Delta E$ & 0.1167 & 0.1157 & 0.128 & 0.121 \\
Al(sc) & $a$ & 2.678 & 2.678 & 2.688 & 2.675\\
& $B$ & 54 & 54 & 57.9 & 60.8\\
& $\Delta E$ & 0.5819 & 0.6018 & 0.479 & 0.483 \\
Al(hcp) & $a$ & 2.806 & 2.805 & 2.846 & 2.793 \\
& $c$ & 1.638 & 1.639 & 1.63 & 1.67\\
& $B$ & 82 & 83 & 71.6 & 80.1\\
& $\Delta E$ & 0.0836 & 0.0821 & 0.032 & 0.038\\
\hline  \hline
\end{tabular}
\end{table}
%
\clearpage
\newpage
\begin{table}
\caption{Comparison of the calculated lattice constants $a$ ($\AA$) and bulk
modulus $B$ (GPa) using atomic orbitals to those from plane wave calculations
for MgO and Pb.
The data of ``Other DZP'', ``Other PW'' and experimental values are taken from
Ref. \onlinecite{junquera01}.
}\label{tab:other}
\begin{tabular}{ccccccc}
\hline \hline
Compound & Properties & PW & DZP & Other DZP & Other PW & Experiments\\
\hline
MgO(B1) & $a$ & 4.233 & 4.238 & 4.11 & 4.10 & 4.21\\
& $B$ & 169 & 169 & 167 & 168 & 152\\
Pb(fcc) & $a$ & 4.873 & 4.875 & 4.88 & 4.88 & 4.95\\
& $B$ & 56 & 55 & 64 & 54 & 43\\
\hline  \hline
\end{tabular}
\end{table}

\clearpage
\newpage

\clearpage
\newpage

\clearpage
\newpage

\clearpage
\newpage

\clearpage
\newpage

\clearpage
\newpage
\clearpage
\newpage

\clearpage
\newpage

\end{document}